# Cognitive States of Potentiality in Art-making

**Nicole Carbert (nicole.carbert@alumni.ubc.ca),**
**Liane Gabora (liane.gabora@ubc.ca),**
**Jasmine Schwartz (jasmineschwartz@hotmail.com),**
and
**Apara Ranjan (apara.ranjan@ubc.ca)**
Department of Psychology, University of British Columbia
Okanagan Campus, 1147 Research Road
Kelowna BC, V1V 1V7
CANADA

**Abstract**

Creativity is thought to involve searching and selecting amongst multiple discrete idea candidates. Honing theory predicts that it involves actualizing the potentiality of as few as a single ill-defined idea by viewing it from different contexts. This paper reports on a study that tests between these theories. Participants were invited to "Create a painting that expresses yourself in any style that appeals to you," and asked "Were all of your ideas for your painting distinct and separate ideas?" Naïve judges were provided with descriptions of the two theories of creativity, sample answers, and practice responses to classify. The judges were significantly more likely to classify the artists' responses as 'H', indicative of honing theory rather than 'S' indicative of a search/select view of creativity.

## Introduction

Creative processes are commonly thought to involve search through a space of possibilities, or the generation of multiple discrete, well-defined candidate ideas followed by selection and exploration of the most promising of them (Runco, 2006). It might seem self-evident that this is the case; how could it be otherwise? However, a different view of creativity is suggested by work on a theory of concepts designed to tackle the difficult problem of modeling how concepts combine. The theory that emerged out of this work, honing theory predicts that creativity involves the merging and interference of memory items resulting in as few as a *single* cognitive structure that is *ill-defined*, and can be said to exist in a state of potentiality, and which can be formally described as a superposition state. The idea becomes increasingly well-defined, and transforms from potential to actual through interaction with both internally generated and externally generated contexts. The idea could actualize in different ways depending on the contexts the idea interacts with, or perspectives it is viewed from. Elements of the seed idea have the potential to grow organically out of earlier elements. In short, honing theory posits that creative ideation involves actualizing the potentiality of as few as a single ill-defined idea by viewing it from different contexts.

These different views of creativity make very different predictions about the state of an idea mid-way through a creative process. This paper reports on a study designed to test between these theories by asking individuals engaged in an art-making task about their conception of their painting mid-way through their creative process.

## Background

Inspired by the metaphor of the mind as a computer (or computer program), early research on creativity focused on the notion of heuristic search, in which rules of thumb guide the inspection of different states within a particular state space (set of possible solutions) until a satisfactory solution is found (Eysenck, 1993; Newell, Shaw & Simon, 1957; Newell & Simon, 1972). In heuristic search, the relevant variables are defined up front; thus the state space is generally fixed. Examples of heuristics include breaking the problem into sub-problems, hill-climbing (reiteratively modifying the current state to look more like the goal state), and working backward from the goal state to the initial state.

The idea that creativity could be construed as heuristically guided search gave hope to those who sought a scientific understanding of creativity because search is a formally tractable process. However it was soon recognized that in many creative tasks, and particularly artistic forms of creativity, the goal state is unspecified, and some elements of the eventual solution may not be present when the problem presents itself. It has been suggested that creativity involves heuristics that guide the search for, not a possibility within a given state space, but a new state space itself (*e.g.*, Boden, 1990; Kaplan & Simon, 1990, Ohlsson, 1992).

One of the most well known theories of creativity is the *Geneplore model* (Finke, Ward, & Smith, 1992). This theory posits that the creative process consists of two stages: generate and explore. (Indeed the name 'Geneplore' is a condensation of 'generate' and 'explore'.) The first stage involves the generation of crudely formed ideas referred to as *pre-inventive structures* that contain the kernel of an idea as opposed to an idea in its entirety. The exploration stage involves fleshing out these pre-inventive structures through elaboration and testing. Use of the term 'exploration' to refer to the second phase of the creative process can be misleading because the term 'explore' is often used to refer to surveying the space of possibilities as generally occurs during the first phase of the creative process, as opposed to refining a single possibility as generally occurs during the second phase. However, the notion of a pre-inventive structure does capture the intuition that early on in the creative process one is working with cognitive structures that are different in kind from those one is working with later in the creative process. The Geneplore model does not attempt to formalize how a pre-inventive structure differs from a full-fledged idea, nor what differentiates a promising pre-inventive structure from a mundane one.

Another well-known theory of creativity is the *Darwinian theory of creativity* (Campbell 1960; Simonton, 1999, 2007, 2010). As do biological species, creative ideas exhibit the kind of complexity and adaptation over time that is indicative of an evolutionary process, not just when they are expressed to others, but in the mind of a single creator (Gabora, 1997; Terrell, Hunt & Gosden, 1997; Thagard, 1980; Tomasello 1996). Thus it has been proposed that in creativity, as in natural selection, there is a process conducive to generating variety, and another conducive to pruning out inferior variants. According to the Darwinian theory, we generate new ideas through an essentially a trial-and-error process involving *'Blind'* generation of ideational *Variants* followed by *Selective Retention* of the fittest variants for development into a finished product. Thus the Darwinian theory is sometimes referred to as BVSR. The variants are said to be blind in the sense that the creator has no subjective certainty about whether they are a step in the direction of the final creative product.

A different view of creativity is suggested by work on a "quantum" theory of concepts designed to tackle the difficult problem of modeling how concepts combine (Aerts, 2009; Aerts, Aerts, & Gabora, 2009; Aerts, Broekaert, Gabora, & Veloz, 2012; Aerts, Gabora, & Sozzo, 2013; Busemeyer, & Bruza, 2012; Gabora, & Aerts, 2002). This theory was developed to cope with the contextuality and noncompositionality of concepts, which make them resistant to formal description. The mathematical formalism of quantum mechanics was developed to describe phenomena that were first observed in the quantum world but that appear also in psychology. One advantage of a quantum model over a classical one is that it uses variables and spaces that are defined specifically with respect to a particular context. The quantum formalism provides a natural spatial representation of a state in a context such that variables are natively context specific.

This theory of concepts led to an alternative view of how the creative process works (Gabora, 2005). Honing theory purports that initially a creative idea exists in an ill-defined state of potentiality that can be formally described as a superposition state. This "seed idea" is experienced as vague because it consists of components potentially derived from multiple sources merged together, and the challenge is to hone it into a well-defined whole. The seed idea transforms from potential to actual through interaction with both internally generated and externally generated contexts. This theory proposes that the idea could actualize in different ways, depending on the contexts the idea interacts with, and that the elements of the idea, have the potential to grow organically out of earlier elements. This theory was in part inspired by extensive discussion with creative individuals, who invariably describe their creative process as involving thinking about an idea from different perspectives as opposed to search and selecting amongst candidate ideas.

The hypothesis that creative thinking involves, not generating and selecting amongst multiple well-formed candidates, but superposition of relevant memory items to give as few as a single candidate that undergoes honing, was tested in an analogy making task (Gabora & Saab, 2011). The 'search and select' view is assumed in the structure mapping (SM) theory of analogy (Gentner, 1983), according to which (in brief) analogy generation occurs in two steps: first, searching memory in a "structurally blind" manner (Gentner, 2010, p. 753) for an appropriate source and aligning it with the target, and second, mapping the correct one-to-one correspondences between the source and the target. Thus, structure mapping assumes that candidate sources are considered separately, and once the correct source is found the analogy making process occurs in isolation from the rest of the contents of mind. This stands in contrast to honing theory in several respects. First, quite the opposite of being "structurally blind", the initial stage of analogy making is thought to be constrained by the content-addressable structure of associative memory to retrieve items that are in some way (although not necessarily the right or most relevant way) structurally similar (Gabora, 2010). Second, honing theory suggests that alignment and mapping may work with, not discrete, predefined structures, but an amalgam of multiple items previously encoded to the neural cell assemblies activated by the target, which in the present context are not readily separated. Third, the process is thought to proceed not by mapping more correspondences but by weeding out non-correspondences.

Participants in the analogy problem solving study were interrupted midway through solving an analogy problem and asked what they were thinking in terms of a solution. Naïve judges categorized a response as H if it met the predictions of honing theory, i.e., if there was evidence of merging solution sources from memory resulting in an ill-defined idea. They categorized it as SM if it met the predictions of structure mapping, i.e., if participants had not finished mapping relations from source to target. Both the frequency counts and mean number of SM versus H judgments supported the hypothesis derived from honing theory that midway through creative processing an idea is in a potentiality state.

This was a first source of empirical evidence that creative thinking is divergent not in the sense that it moves in multiple directions or generates multiple possibilities, but in the sense that it produces a raw idea that is vague or unfocused, that requires further processing to become viable. However, analogy problem solving is a highly

Carbert, N., Gabora, L., Schwartz, J., & Ranjan, A. (2014). States of cognitive potentiality in art-making. *Proceedings of the AIEA Congress on Empirical Aesthetics* (pp. 121-126). Held August 22-24, New York.constrained task, with a single correct solution, and one could argue that this is uncharacteristic of most creative tasks. The goal of the current research is to show that the honing theory of creativity is broader than just analogy problem solving, that it applies to less constrained, open-ended task such as art making. This is the first study to investigate between the above two views of how the creative processes works using an open-ended task. We hypothesized that midway through the process of completing a painting the mind of the artist is in a state of potentiality, i.e*.,* the idea the artist has is vague or half-baked, and in different contexts it could manifest different ways. We propose that the creative process works through honing an idea such that it transforms from ill-defined to well-defined, as opposed to search and selection from amongst a collection of well-formed candidate ideas. Attempting to distinguish between these competing theories of creativity helps us to understand how creative ideas arise and take shape, and thus enables us to gain insight into human innovation.

## Method

The study consisted of two components, carried out separately. The first involved the generation of artworks and the answering of questions concerning the art-making process. The second involved classifying these responses in ways that were indicative of either honing theory or a search and select type theory.

### Participants

The participants were undergraduates enrolled in psychology courses at the University of British Columbia who were participating in order to receive class credit. There were two types of participants: 56 who created paintings and answered questionnaires, who will be referred to as *artists*, and 11 who judged the artists' answers to the questionnaires, who will be referred to as *judges*. All participants were naïve concerning the rationale for the study.

### Protocol for Artists

**Materials** Between two and eight artists participated at a time. The artists' were seated at desks placed in a circle, with each individual facing outwards, such that they could not see each others' art. Each artist's desk had paintbrushes of two different of sizes and an ice cube tray containing seven different colors (pink, white, yellow, green, brown, blue, and red) of acrylic paint. Each desk also had a glass of water and towelettes for cleaning the paintbrushes, and a plastic plate for mixing colors. Additionally, on the desk was a set of ten Crayola washable watercolor paints in plastic jars, pencil crayon pastels, chalk pastels, oil pastels, and two pieces of paper towel, a piece of watercolor paper, and a piece of paper for acrylic paints.

**Handouts and Art-making Protocol** Each artist was also provided with a set of handouts. The first was a consent form, and the second was a Demographics Questionnaire. **Form A** asked them to "Create a painting that expresses yourself in any style that appeals to you." The experimenter told them that their artwork would not be judged in any way, and that it was only their answers to questions on Forms B, C, and D that would be analyzed. They were told that the study was concerned with peoples' perceptions of the creative process of art making. They were told where and when they could pick up their painting if they wished to, and that if they did not pick it up it would be destroyed after one year. They were also told that none of the artwork would be photographed.

Before beginning the painting they were asked to fill in **Form B**, which asked:

> What are you thinking in terms of what your painting will look like? Write down your thoughts about your painting in as much detail as you can.

They were asked to start painting as soon as they finished writing their answer to the question on Form B.

After 15 minutes the artists were interrupted and given **Form C,** which asked them to:

> Write down your thoughts about the painting in as much detail as you can.

After they finished responding to Form C they were asked to continue their painting until it was finished.

Once they had completed their painting they were given **Form D**, which asked:

> Were all of your ideas for your painting distinct and separate ideas?

In a Pilot Study with 15 participants there was an additional question on Form D that asked: Do you think being interrupted affected how you carried out your painting? Since all of the participants answered "no" to this question it was thereafter omitted. (Had any of them answered "yes" we would have omitted Form C.)

When the artists completed Form D they were debriefed and asked if they had any questions regarding the study. They all completed the procedure within ninety minutes.

### Judging

The judges did not see or evaluate any of the actual artworks; all they looked at was the artists' questionnaire responses. Each judge was provided with a set of handouts. As with the artists, the first two were a consent form and a demographics questionnaire. The third handout described the criteria they were to use to classify artist answers as indicating of one theory or the other. It read as follows:

> **JUDGING CRITERIA**
> You will be trained how to read information about artists' art-making processes and then put them into two categories: one that you think is indicative of one theory about the creative art-making process, and another that is indicative of another theory. You will first be given the identifying characteristics of Theory S or Theory H. You will practice on toy examples until your answers indicate that you understand the distinction between H and S, and until you yourself claim that you understand this distinction. Then you will classify the real responses as S or H.

**Identifying Characteristics of Theory S**

- If multiple ideas are given, they are considered separate and distinct from one another; not hard to dis-entangle.
- Does not contain extra ideas that would be relevant to other types of paintings or creative tasks but that are irrelevant to this particular painting.
- Distinct possible ways of going about the task may be separated by words such as "or" without anything to indicate these two ways are connected in creator's mind.
- The creative process involves searching one's mind for ways to go about the painting and selecting amongst these distinct possible outcomes.
- No common core to possible painting outcomes.
- No new emergent characteristics of painting come to light through process of resolving how initial idea for painting will be carried out.

**Identifying Characteristics of Theory H**

- If multiple ideas are given, they are jumbled together; hard to dis-entangle.
- Contains extra ideas that would be relevant to other types of creative tasks but that is not relevant to creating this particular painting.
- Ill-defined or indistinct ideas; challenge is to make them concrete.
- Common core to different possible painting outcomes; core could be taken in different directions.
- Words indicative of H theory: vague, ill-defined, indistinct, potential.
- New emergent characteristics of painting, or new self-understanding, come to light through process of resolving how initial idea for painting will be carried out; transformative.

Note: It is not necessary to meet *all* the criteria in order to be properly classified as indicative of one theory of the other. Also, it is not necessary to distribute your classifications evenly; it is possible that 5 in a row might be indicative of one theory and not the other. In fact it is completely possible that they all could be one theory and not the other. Please classify each answer as indicative of one theory or the other even if you are not sure of your answer.

The Judging Criteria Handout also contained a summary of differences between the two theories provided in Table 1. The judges were told that a flat-out 'yes' to 'Were all of your ideas for your painting distinct and separate ideas?' was indicative of S and a flat-out 'no' was indicative of H.

The judges were given the 56 questionnaires, and a form on which to classify the artists' answers as indicative of either H or S. They were told "If you are not sure just go with your gut feeling".

Table 1: Characteristics used to judge responses as Search and Select (S) versus Honing (H).

|  | S | H |
|---|---|---|
| If multiple ideas are given, they are | Distinct (e.g., complete ideas separated by 'or') | Jumbled together (e.g., idea fragments spliced together) |
| Ideas are | Well-defined; need to be tweaked / mutated and selected amongst | Ill-defined; need to be made concrete; later elements emerge from earlier ones |
| Common core to ideas? | Never | Yes or sometimes |
| Emergent properties of painting? | No | Yes |
| Emergent self-understanding? | No | Yes |
| Emphasis | External product | Internal transformation |

## Results

Not all artists answered every question, and not all judges provided a judgment of every artist response. The intra class correlation (ICC) for the degree of agreement amongst the judges' classifications of the responses for Forms B and C were too low to use and so were excluded from the analysis. For the question on Form D the ICC was .94, indicating high reliability.

Artist responses were classified as supporting honing theory if 6 or more judges judged it as H, and as supporting search and select if 6 or more judges judged it as S. As shown in Figure 1, 29 of the artist responses were classified by the judges as supporting of theory H, and 13 were classified as supporting theory S. A one-sample chi-square test revealed a statistically significant difference between the classifications for theory H and S, $\chi^2(1, N = 43) = 4.57$, $p < .05$. Thus the frequency counts support the hypothesis that the creative process involves actualization of the potential of as few as one single ill-defined idea, as predicted by honing theory, as opposed to searching and selecting from amongst multiple discrete variants, as predicted by theories such as BVSR.

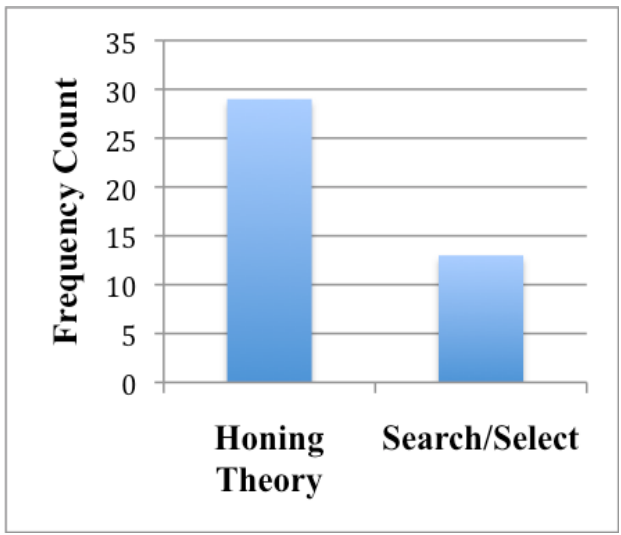

Figure 1: Frequency count of judgments for Honing Theory (H), on left (*N* = 29) and Search and Select Theory (S), on right. (*N* = 13).

A further analysis compared the mean number of judgments (out of a maximum of 11, the total number of judges) across all responses that supported each theory. Taking the mean across all 56 responses, the mean number of S judgments was 9.67 (SD = 0.47), and the mean number of H judgments was 4.35 (SD = 0.47). A paired-sample *t*-test showed that the difference was significant $t(55) = 2.64$, $p < .05$), and the effect size ($\eta^2 = .38$) was large. Thus these data corroborate the above frequency count findings. The mean judgment scores for structure mapping and actualizing potentiality are given in Figure 2.

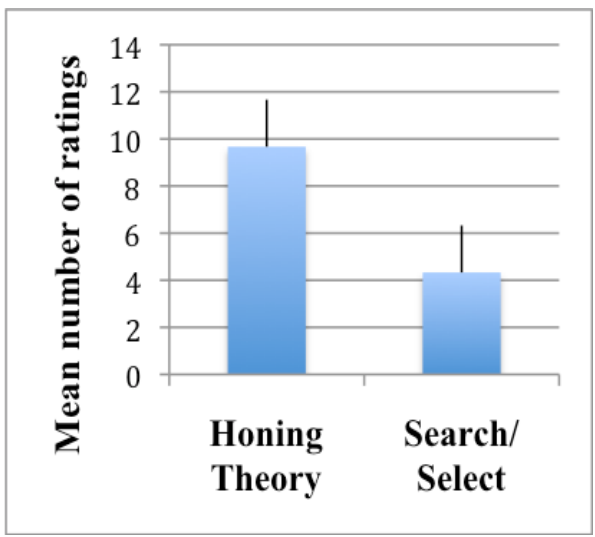

Figure 2: Mean number of ratings of H (indicative of honing theory) on the left and S (indicative of a search and select view of creativity) on the right (*M* = 4.33, SD = 0.47).

## Discussion

This study investigated whether the mind of the artist is in a state of potentiality midway through the open-ended creative task of art making, helping us to better understand how the creative process works. Our goal was to determine if creative ideas arise through the generation of multiple, discrete, well-defined possibilities that are then explored, tweaked, and selected amongst, or if creative ideas arise through the merging of memory items resulting in a single cognitive structure that is ill defined, thus existing in a state of potentiality that becomes more well-defined as it interacts with internally or externally generated contexts. Our results support the hypothesis that the creative process works through honing—actualization of potentiality—as opposed to search and selection. Although this hypothesis had previously found support with respect to analogy problem solving (Gabora & Saab, 2011), this is the first study to investigate and support that it also holds true for this open-ended creative task of art making.

The data collected for the study reported here was part of an Honours Student project and it was carried out with the view of ironing out the kinks in the procedure and thereby paving the way for a larger study of this sort in the future. There are several limitations to the research reported here. First a larger subject pool is required; perhaps due in part to the creatively demanding nature of the task it, was difficult to get enough participants to obtain statistically significant results. Second, in future studies that use this protocol, more time should be spent with the judges to ensure that they can consistently judge the more open-ended questions (Forms B and C) as well as the more clear-cut question on Form D. Future research should also investigate the role of potentiality states in other types of creative tasks. Another aspect of the study that could be improved is the questionnaires. Future studies should expand on the types of questions and carry out extensive pilot studies to assess their inter-rater reliability in order to tap into the cognitive processes during a creative task more adequately. In addition, the questionnaires could be complimented by behavioral observations of the artsts at work so as to not be relying completely on self- perceptions of their creative process. Nevertheless we view the results as a promising step forward to empirically investigating the cognitive process by which creative works come into being.

## Acknowledgments

Funding for this research comes from the Natural Sciences and Engineering Research Council of Canada and the Flemish Fund for Scientific Research.